\documentclass[prl,twocolumn,superscriptaddress,showkeys]{revtex4}
\usepackage{amssymb,epsf,epsfig}

\newcommand{\be}{\begin{equation}}
\newcommand{\ee}{\end{equation}}
\newcommand{\ord}{{\cal O}}
\newcommand{\lsim}{
\mathrel{\hbox{\rlap{\hbox{\lower4pt\hbox{$\sim$}}}\hbox{$<$}}}}
\newcommand{\gsim}{
\mathrel{\hbox{\rlap{\hbox{\lower4pt\hbox{$\sim$}}}\hbox{$>$}}}}

\def\kpn{K^+\rightarrow\pi^+\nu\bar\nu}
\def\klpn{K_{\rm L}\rightarrow\pi^0\nu\bar\nu}
\def\epe{\varepsilon'/\varepsilon}

\begin{document}
\begin{titlepage}
\vspace*{0.7truecm}
\begin{flushright}
CERN-TH/2003-314\\
TUM-HEP-538/03\\
MPP-2003-142\\
hep-ph/0312259
\end{flushright}

\vspace{1.6truecm}
\begin{center}
\boldmath
{\Large\bf $B\to\pi\pi$, New Physics in $B\to\pi K$ and Implications for

\vspace*{0.3truecm}

Rare $K$ and $B$ Decays}
\unboldmath
\end{center}

\vspace{1.2truecm}

\begin{center}
{\bf \large Andrzej J. Buras,${}^a$ Robert Fleischer,${}^b$ 
Stefan Recksiegel${}^a$ and Felix Schwab${}^{c,a}$}

\vspace{0.7truecm}

${}^a$ {\sl Physik Department, Technische Universit\"at M\"unchen,
D-85748 Garching, Germany}

\vspace{0.2truecm}

${}^b$ {\sl Theory Division, CERN, CH-1211 Geneva 23, Switzerland}

\vspace{0.2truecm}

 ${}^c$ {\sl Max-Planck-Institut f{\"u}r Physik -- Werner-Heisenberg-Institut,
 D-80805 Munich, Germany}

\end{center}

\vspace*{1.7cm}

\begin{center}
\large{\bf Abstract}

\vspace*{0.6truecm}

\begin{tabular}{p{14.5truecm}}
{\small 
The measured $B\to\pi\pi,\pi K$ branching ratios exhibit puzzling patterns. We 
point out that the $B\to\pi\pi$ hierarchy can be nicely accommodated in the 
Standard Model (SM) through non-factorizable hadronic interference effects,
whereas the $B\to\pi K$ system may indicate new physics (NP) in the electroweak
(EW) penguin sector. Using the $B\to\pi\pi$ data and the $SU(3)$ flavour 
symmetry, we may fix the hadronic $B\to\pi K$ parameters, which allows us to 
show that any currently observed feature of the $B\to\pi K$ system can be 
easily explained through enhanced EW penguins with a large CP-violating NP 
phase. Restricting ourselves to a specific scenario, where NP enters only 
through $Z^0$ penguins, we derive links to rare $K$ and $B$ decays, where an 
enhancement of the $K_{\rm L}\to\pi^0\nu\bar\nu$ rate by one order of 
magnitude, with $\mbox{BR}(\klpn)>\mbox{BR}(\kpn)$, 
$\mbox{BR}(K_{\rm L}\to\pi^0 e^+e^-)=\ord(10^{-10})$, 
$(\sin2\beta)_{\pi\nu\bar\nu}<0$, and a large forward--backward CP asymmetry 
in $B_d\to K^*\mu^+\mu^-$, are the most spectacular effects. We address 
also other rare $K$ and $B$ decays, $\epe$ and $B_d\to \phi K_{\rm S}$.}
\end{tabular}

\end{center}

\vspace*{1.7truecm}

\noindent
December 2003

\end{titlepage}

\newpage
\thispagestyle{empty}
\mbox{}

\newpage
\thispagestyle{empty}
\mbox{}

\rule{0cm}{23cm}

\newpage
\thispagestyle{empty}
\mbox{}

\setcounter{page}{0}

\preprint{hep-ph/0312259}
\preprint{CERN-TH/2003-314}
\preprint{TUM-HEP-538/03}
\preprint{MPP-2003-142}
\date{December 18, 2003}

\title{\boldmath$B\to\pi\pi$, New Physics in $B\to\pi K$ and 
Implications for Rare $K$ and $B$ Decays\unboldmath}

\author{Andrzej J. Buras}
%\email{Andrzej.Buras@ph.tum.de}
\affiliation{Physik Department, Technische Universit\"at M\"unchen,
D-85748 Garching, Germany}

\author{Robert Fleischer}
%\email{Robert.Fleischer@cern.ch}
\affiliation{Theory Division, CERN, CH-1211 Geneva 23, Switzerland}

\author{Stefan Recksiegel}
%\email{Stefan.Recksiegel@ph.tum.de}
\affiliation{Physik Department, Technische Universit\"at M\"unchen,
D-85748 Garching, Germany}

\author{Felix Schwab}
%\email{Felix.Schwab@ph.tum.de}
\affiliation{Max-Planck-Institut f{\"u}r Physik -- Werner-Heisenberg-Institut,
D-80805 Munich, Germany}
\affiliation{Physik Department, Technische Universit\"at M\"unchen,
D-85748 Garching, Germany}

\begin{abstract}
\vspace{0.2cm}\noindent
The measured $B\to\pi\pi,\pi K$ branching ratios exhibit puzzling patterns. We 
point out that the $B\to\pi\pi$ hierarchy can be nicely accommodated in the 
Standard Model (SM) through non-factorizable hadronic interference effects,
whereas the $B\to\pi K$ system may indicate new physics (NP) in the electroweak
(EW) penguin sector. Using the $B\to\pi\pi$ data and the $SU(3)$ flavour 
symmetry, we may fix the hadronic $B\to\pi K$ parameters, which allows us to 
show that any currently observed feature of the $B\to\pi K$ system can be 
easily explained through enhanced EW penguins with a large CP-violating NP 
phase. Restricting ourselves to a specific scenario, where NP enters only 
through $Z^0$ penguins, we derive links to rare $K$ and $B$ decays, where an 
enhancement of the $K_{\rm L}\to\pi^0\nu\bar\nu$ rate by one order of 
magnitude, with $\mbox{BR}(\klpn)>\mbox{BR}(\kpn)$, 
$\mbox{BR}(K_{\rm L}\to\pi^0 e^+e^-)=\ord(10^{-10})$, 
$(\sin2\beta)_{\pi\nu\bar\nu}<0$, and a large forward--backward CP asymmetry 
in $B_d\to K^*\mu^+\mu^-$, are the most spectacular effects. We address 
also other rare $K$ and $B$ decays, $\epe$ and $B_d\to \phi K_{\rm S}$.
\end{abstract}

\keywords{Non-leptonic $B$ decays, CP violation, rare $K$ and $B$ decays} 

\maketitle

{\bf 1.}\ In this letter, we consider simultaneously the decays 
$B\to\pi\pi$, $B\to\pi K$ and prominent rare $K$ and $B$ decays 
within the SM and its simple extension in which NP enters dominantly 
through enhanced EW penguins with new weak phases. Our analysis consists 
of three interrelated parts, and has the following logical structure:

i) Since $B\to\pi\pi$ decays and the usual analysis of the
unitarity triangle (UT) are only insignificantly affected by EW penguins,
the $B\to\pi\pi$ system can be described as in the SM and allows the 
extraction of the relevant hadronic parameters by assuming only the 
isospin symmetry. The values of these parameters imply important
non-factorizable contributions, and allow us to predict the CP-violating
$B_d\to\pi^0\pi^0$ observables. 

ii) Using the $SU(3)$ flavour symmetry and plausible dynamical assumptions,
we may determine the hadronic $B\to\pi K$ parameters through their 
$B\to\pi\pi$ counterparts, and may analyse the $B\to\pi K$ system in 
the SM. Interestingly, those observables where EW penguins play a minor 
r\^ole are found to agree with the pattern of the $B$-factory data. On the 
other hand, the observables that are significantly affected by EW penguins 
are found to disagree with the experimental picture, thereby suggesting NP 
in the EW penguin sector. Indeed, we may describe all the currently available 
data through sizeably enhanced EW penguins with a large CP-violating NP 
phase around $-90^\circ$, and may then predict the CP-violating 
$B_d\to \pi^0 K_{\rm S}$ observables. Moreover, we may obtain insights 
into $SU(3)$-breaking effects, which support our working assumptions, 
and may determine the UT angle $\gamma$, in accordance with the well-known 
UT fits.  

iii) In turn, the enhanced EW penguins, with their large CP-violating NP 
phases, have important implications for rare $K$ and $B$ decays, with 
several predictions that are significantly different from the SM 
expectations.

This letter summarizes the most interesting results of each step. 
The details behind the findings presented here are described in 
\cite{BFRS-III}, where the arguments for the assumptions made in our 
analysis are spelled out, other results are presented, and a detailed
list of references is given. 

{\bf 2.}\
The BaBar and Belle collaborations have very recently reported the observation 
of $B_d\to\pi^0\pi^0$ decays with CP-averaged branching ratios of 
$(2.1\pm0.6\pm0.3)\times10^{-6}$ and $(1.7\pm0.6\pm0.2)\times10^{-6}$, 
respectively \cite{Babar-Bpi0pi0,Belle-Bpi0pi0}. These measurements represent
quite a challenge for theory. For example, in a recent state-of-the-art 
calculation within QCD factorization \cite{Be-Ne}, a branching ratio that 
is about six times smaller is favoured, whereas the calculation of
$B_d\to\pi^+\pi^-$ points towards a branching ratio about two times 
larger than the current experimental average. On the other hand, the
calculation of $B^+\to\pi^+\pi^0$ reproduces the data rather well. This
``$B\to\pi\pi$ puzzle'' is reflected by the following quantities:
\begin{eqnarray}
R_{+-}^{\pi\pi}&\equiv&2\left[\frac{\mbox{BR}(B^\pm\to\pi^\pm\pi^0)}{\mbox{BR}
(B_d\to\pi^+\pi^-)}\right]\frac{\tau_{B^0_d}}{\tau_{B^+}}=
2.12\pm0.37~~\mbox{}\label{Rpm-def}\\
R_{00}^{\pi\pi}&\equiv&2\left[\frac{\mbox{BR}(B_d\to\pi^0\pi^0)}{\mbox{BR}
(B_d\to\pi^+\pi^-)}\right]=0.83\pm0.23,\label{R00-def}
\end{eqnarray}
where we have used $\tau_{B^+}/\tau_{B^0_d}=1.086\pm0.017$ and 
the most recent compilation of the Heavy Flavour Averaging Group (HFAG) 
\cite{HFAG}; the central values calculated within QCD factorization 
\cite{Be-Ne} give $R_{+-}^{\pi\pi}=1.24$ and $R_{00}^{\pi\pi}=0.07$.
In order to simplify our $B\to\pi\pi$ analysis, we neglect EW penguins, 
which play a minor r\^ole in these transitions and can be straightforwardly 
included through the isospin symmetry \cite{BFRS-III,BF-neutral1,GPY}. 
We then have 
\begin{eqnarray}
\sqrt{2}A(B^+\to\pi^+\pi^0)&=&-[\tilde T+\tilde C]\label{B+pi+pi0}\\
A(B^0_d\to\pi^+\pi^-)&=&-[\tilde T + P]\label{Bdpi+pi-}\\
\sqrt{2}A(B^0_d\to\pi^0\pi^0)&=&-[\tilde C - P],\label{Bdpi0pi0}
\end{eqnarray}
where 
\begin{eqnarray}
P&=&\lambda^3 A({\cal P}_t-{\cal P}_c)\equiv
\lambda^3 A{\cal P}_{tc}\label{P-def}\\
\tilde T &=&\lambda^3 A R_b e^{i\gamma}\left[{\cal T}-\left({\cal P}_{tu}-
{\cal E}\right)\right]\label{T-tilde}\\
\tilde C &=&\lambda^3 A R_b e^{i\gamma}\left[{\cal C}+\left({\cal P}_{tu}-
{\cal E}\right)\right].\label{C-tilde}
\end{eqnarray}
Here $\lambda$, $A$ and $R_b\propto |V_{ub}/V_{cb}|$ parametrize
the Cabibbo--Kobayashi--Maskawa (CKM) matrix, the ${\cal P}_q$ are the 
strong amplitudes of QCD penguins with internal $q$-quark exchanges 
($q\in\{t,c,u\}$), including annihilation and exchange penguins, while 
${\cal T}$ and ${\cal C}$ are the strong amplitudes of colour-allowed and 
colour-suppressed tree-diagram-like topologies, respectively, and 
${\cal E}$ denotes exchange topologies. Introducing the hadronic parameters
\begin{eqnarray}
d e^{i\theta}&\equiv&-e^{i\gamma}P/{\tilde T}=
-|P/{\tilde T}|e^{i(\delta_P-\delta_{\tilde T})}\\
x e^{i\Delta}&\equiv&{\tilde C}/{\tilde T}=
|{\tilde C}/{\tilde T}|e^{i(\delta_{\tilde C}-
\delta_{\tilde T})},\label{x-Bpipi}
\end{eqnarray}
with the strong phases $\delta_P$, $\delta_{\tilde T}$ and 
$\delta_{\tilde C}$, we obtain
\begin{equation}\label{Rpm-res}
R_{+-}^{\pi\pi}=\frac{1+2x\cos\Delta+x^2}{1-
2d\cos\theta\cos\gamma+d^2}
\end{equation}
\vspace*{-0.4truecm}
\begin{equation}\label{R00-res}
R_{00}^{\pi\pi}=\frac{d^2+2dx\cos(\Delta-\theta)\cos\gamma+
x^2}{1-2d\cos\theta\cos\gamma+d^2}
\end{equation}
\vspace*{-0.4truecm}
\begin{equation}\label{Adir-Bpipi}
{\cal A}_{\rm CP}^{\rm dir}=-\left[\frac{2d\sin\theta\sin\gamma}{1-
2d\cos\theta\cos\gamma+d^2}\right]
\end{equation}
\vspace*{-0.4truecm}
\begin{equation}\label{Amix-Bpipi}
{\cal A}_{\rm CP}^{\rm mix}=\frac{\sin(\phi_d+2\gamma)-2d\cos\theta
\sin(\phi_d+\gamma)+d^2\sin\phi_d}{1-2d\cos\theta\cos\gamma+d^2},
\end{equation}
where $\phi_d$ denotes the $B^0_d$--$\bar B^0_d$ mixing phase
and ${\cal A}_{\rm CP}^{\rm dir}$ and ${\cal A}_{\rm CP}^{\rm mix}$
are the direct and mixing-induced $B_d\to\pi^+\pi^-$ CP asymmetries 
\cite{RF-BsKK,Fl-Ma}. The currently available BaBar 
\cite{BaBar-Bpipi} and Belle \cite{Belle-Bpipi} results for 
${\cal A}_{\rm CP}^{\rm dir}(\pi^+\pi^-)$ and 
${\cal A}_{\rm CP}^{\rm mix}(\pi^+\pi^-)$ are not fully consistent 
with each other. If one calculates, nevertheless, the 
weighted averages, one finds \cite{HFAG} 
\be\label{AdirBpipi-exp}
{\cal A}_{\rm CP}^{\rm dir}(\pi^+\pi^-)=-0.38\pm0.16,\quad 
{\cal A}_{\rm CP}^{\rm mix}(\pi^+\pi^-)=0.58\pm0.20. 
\ee

As pointed out in \cite{Fl-Ma,FIM}, in the case of $\phi_d\sim 47^\circ$,
the CP asymmetries in (\ref{AdirBpipi-exp}) point
towards $\gamma\sim 60^\circ$, in accordance  with the SM. In the 
following, our main focus is on the hadronic parameters. If we assume 
that $\gamma=(65\pm 7)^\circ$ and $\phi_d=2\beta=(47\pm4)^\circ$, as in the 
SM \cite{CKM-Book}, (\ref{Rpm-res})--(\ref{Amix-Bpipi}) and the data in
(\ref{Rpm-def}), (\ref{R00-def}), (\ref{AdirBpipi-exp})
imply
\begin{equation}\label{d-x-det}
\begin{array}{rclcrcl}
d&=&0.49^{+0.33}_{-0.21}, &&\theta&=&+\left(137^{+19}_{-23}\right)^\circ,\\
x&=&1.22^{+0.25}_{-0.21}, &&\Delta&=&-\left(71^{+19}_{-25}\right)^\circ,
\end{array}
\end{equation}
where we have suppressed a second solution for $(x,\Delta)$, which 
does not allow us to accommodate the $B\to\pi K$ data \cite{BFRS-III}.
This determination is essentially theoretically clean, and the experimental 
picture will improve significantly in the future. 
We observe that $x={\cal O}(1)$, which implies $|\tilde C|\sim |\tilde T|$. 
In view of the anticipated colour suppression of ${\cal C}$ with respect to
${\cal T}$, this can only be satisfied if the usually neglected
contributions $({\cal P}_{tu}-{\cal E})$ in (\ref{T-tilde}) and 
(\ref{C-tilde}) are significant \cite{PAP0}. Indeed, because of the different 
signs in (\ref{T-tilde}) and (\ref{C-tilde}), we may explain the 
surprisingly small $B_d\to\pi^+\pi^-$ branching ratio naturally, through 
{\it destructive} interference between the ${\cal T}$ and 
$({\cal P}_{tu}-{\cal E})$ amplitudes, whereas the puzzling large 
$B_d\to\pi^0\pi^0$ branching ratio originates from {\it constructive} 
interference between the ${\cal C}$ and $({\cal P}_{tu}-{\cal E})$ 
amplitudes. Within factorization, $B_d\to\pi^+\pi^-$ would favour 
$\gamma>90^\circ$, in contrast to the SM expectation, thereby 
reducing BR$(B_d\to\pi^+\pi^-)$ through destructive interference between trees 
and penguins. Now we arrive at a picture that is very different from 
factorization and exhibits certain interference effects at the hadronic 
level; this allows us to accommodate straightforwardly {\it any} currently 
observed feature of the $B\to\pi\pi$ system within the SM. Moreover, we 
may {\it predict} the CP-violating $B_d\to\pi^0\pi^0$ observables 
\cite{BFRS-III}: 
\begin{equation}
{\cal A}_{\rm CP}^{\rm dir}(\pi^0\pi^0)%(B_d\to\pi^0\pi^0)
=-0.40^{+0.35}_{-0.18},\quad
{\cal A}_{\rm CP}^{\rm mix}(\pi^0\pi^0)%(B_d\to\pi^0\pi^0)
=-0.56^{+0.43}_{-0.44}.
\end{equation}

{\bf 3.}\ 
In the $B\to\pi K$ system, the following ratios of CP-averaged branching 
ratios are of central interest \cite{BF-neutral1}:
\begin{eqnarray}
R_{\rm c}&\equiv&2\left[\frac{\mbox{BR}(B^\pm\to\pi^0K^\pm)}{\mbox{BR}
(B^\pm\to\pi^\pm K^0)}\right]=1.17\pm0.12\label{Rc-def}\\
R_{\rm n}&\equiv&\frac{1}{2}\left[
\frac{\mbox{BR}(B_d\to\pi^\mp K^\pm)}{\mbox{BR}(B_d\to\pi^0K)}\right]=
0.76\pm0.10,\label{Rn-def}
\end{eqnarray}
with numerical values following from \cite{HFAG}. As noted in 
\cite{BF-neutral2}, the pattern of $R_{\rm c}>1$ and $R_{\rm n}<1$ is actually 
very puzzling. On the other hand,
\begin{equation}\label{R-def}
R\equiv\left[\frac{\mbox{BR}(B_d\to\pi^\mp K^\pm)}{\mbox{BR}
(B^\pm\to\pi^\pm K)}\right]\frac{\tau_{B^+}}{\tau_{B^0_d}}
=0.91\pm0.07
\end{equation}
does not show any anomalous behaviour. Since $R_{\rm c}$ and $R_{\rm n}$ are 
affected significantly by colour-allowed EW penguins, whereas these topologies 
may only contribute in colour-suppressed form to $R$, this ``$B\to\pi K$ 
puzzle'' may be a manifestation of NP in the EW penguin sector 
\cite{BF-neutral2,BFRS-I}, offering an attractive avenue for physics beyond 
the SM to enter the $B\to\pi K$ system \cite{trojan}. 

In this letter, we neglect colour-suppressed EW penguins, employ $SU(3)$ 
flavour-symmetry 
arguments, and assume that penguin annihilation and exchange topologies 
are small. The latter topologies can be probed through $B_d\to K^+K^-$,
where the current experimental bound of $\mbox{BR}(B_d\to K^+K^-)<
0.6\times 10^{-6}~(\mbox{90\% C.L.})$ \cite{HFAG} does not indicate 
any anomalous behaviour \cite{BFRS-III}. We then go beyond \cite{BFRS-I} 
in two respects. First, we employ 
the $B\to\pi\pi$ data to fix the hadronic parameters of the $B\to\pi K$ 
system. Second, we consider CP-violating NP contributions to the EW penguin 
sector, so that these topologies are described by a parameter $q$ with a 
CP-violating weak phase $\phi$, which vanishes in the SM. We may then write 
\begin{equation}
A(B^0_d\to\pi^-K^+)=P'\left[1-re^{i\delta}e^{i\gamma}\right]
\end{equation}
\vspace*{-0.8truecm}
\begin{equation}
\sqrt{2}A(B^0_d\to\pi^0K^0)=-P'\left[1+\rho_{\rm n}e^{i\theta_{\rm n}}
e^{i\gamma}-qe^{i\phi}r_{\rm c}e^{i\delta_{\rm c}}\right],
\end{equation}
where $P'\equiv\left(1-\lambda^2/2\right)A\lambda^2({\cal P}_t-{\cal P}_c)$
is the counterpart of (\ref{P-def}), and the $B\to\pi\pi$ analysis described
above allows us to fix the hadronic $B\to\pi K$ parameters through
\cite{BFRS-III}
\begin{eqnarray}
re^{i\delta}&\equiv&\left(\frac{\lambda^2R_b}{1-\lambda^2}
\right)\left[\frac{{\cal T}-({\cal P}_t-{\cal P}_u)}{{\cal P}_t-
{\cal P}_c}\right]=-\frac{\epsilon}{de^{i\theta}}\\
\rho_{\rm n}e^{i\theta_{\rm n}}&\equiv&\left(\frac{\lambda^2R_b}{1-\lambda^2}
\right)\left[\frac{{\cal C}+({\cal P}_t-{\cal P}_u)}{{\cal P}_t-
{\cal P}_c}\right]=xe^{i\Delta}re^{i\delta}\mbox{}~~~\\
r_{\rm c}e^{i\delta_{\rm c}}&\equiv&\left(\frac{\lambda^2R_b}{1-\lambda^2}
\right)\left[\frac{{\cal T}+{\cal C}}{{\cal P}_t-{\cal P}_c}\right]
=re^{i\delta}+\rho_{\rm n}e^{i\theta_{\rm n}},
\end{eqnarray}
where $\epsilon\equiv\lambda^2/(1-\lambda^2)=0.05$.
Consequently, (\ref{d-x-det}) yields
\begin{equation}
\begin{array}{rclcrcl}
r&=&0.11^{+0.06}_{-0.05}, &&  \delta&=&+(43^{+23}_{-19})^\circ,\\
\rho_{\rm n}&=&0.13^{+0.06}_{-0.05},  && 
\theta_{\rm n}&=&-(28^{+21}_{-26})^\circ,\\
r_{\rm c}&=&0.20^{+0.09}_{-0.07}, && \quad  
\delta_{\rm c}&=&+(3^{+23}_{-18})^\circ,
\end{array}
\end{equation}
where the errors have been added in quadrature.
We observe that $re^{i\delta}$ and $\rho_{\rm n}e^{i\theta_{\rm n}}$ differ 
strongly from factorization. However, the small value of $r$ implies 
generically small CP violation in $B_d\to\pi^\mp K^\pm$ at the $10\%$
level \cite{BFRS-III}, in accordance with the data \cite{HFAG}. 
Interestingly, the value of $r_{\rm c}$ agrees well with the one of an 
alternative determination through $B^\pm\to\pi^\pm\pi^0, \pi^\pm K$ 
decays \cite{GRL}, $0.196\pm0.016$, thereby pointing towards moderate 
non-factorizable $SU(3)$-breaking corrections. 
The charged $B\to\pi K$ modes involve an additional parameter 
$\rho_{\rm c}e^{i\theta_{\rm c}}\propto\lambda^2R_b$ 
\cite{BFRS-III}, which is estimated to contribute at the few-per cent level,
and is neglected, as in \cite{BFRS-I}. This assumption is also supported
by the searches for direct CP violation in $B^\pm\to\pi^\pm K$ and 
the experimental upper bounds for $\mbox{BR}(B^\pm\to K^\pm K)$ \cite{HFAG}. 
We may then write
\begin{equation}
A(B^+\to\pi^+K^0)=-P'
\end{equation}
\vspace*{-0.8truecm}
\begin{equation}
\sqrt{2}A(B^+ \!\to\! \pi^0K^+)=P'\left[1 \!-\! \left(e^{i\gamma} \!-\!
qe^{i\phi}\right)r_{\rm c}e^{i\delta_{\rm c}}\right],
\end{equation}
allowing us to study the $R_{\rm c,n}$ and the relevant $B\to\pi K$ CP 
asymmetries as functions of $q$ and $\phi$. We find -- in accordance with 
\cite{BFRS-I} -- that the data in (\ref{Rc-def}) and (\ref{Rn-def}) cannot 
be described properly for the SM values $q=0.69$ \cite{NR} and $\phi=0$;
in particular, $R_{\rm c}\sim 1.14$ and $R_{\rm n}\sim 1.11$. 
However, treating $q$ and 
$\phi$ as free parameters, we obtain
\begin{equation}\label{q-det}
q=1.78^{+1.24}_{-0.97}, \quad \phi=-(85^{+11}_{-13})^\circ,
\end{equation}
and a generically small CP asymmetry in $B^\pm \to \pi^0K^\pm$,
in accordance with the data \cite{BFRS-III}. 
If we allow for a strong phase $\omega$ in the EW penguin sector, which 
may be induced by non-factorizable $SU(3)$-breaking effects
\cite{BF-neutral1}, the data for this CP asymmetry and the $R_{\rm c,n}$ 
allow us to determine $\omega$ as well. We find a phase at the few-degree 
level and essentially unchanged values of $q$ and $\phi$, thereby giving 
us additional support for the use of the $SU(3)$ flavour symmetry 
\cite{BFRS-III}. 
In contrast to \cite{BFRS-I}, where larger direct CP asymmetries in the 
$B\to\pi K$ modes were favoured, the determination of the hadronic parameters 
through the $B\to\pi\pi$ system and the introduction of the 
weak EW penguin phase $\phi$ now allow us to describe {\it any} 
currently observed feature of the $B\to\pi K$ modes. Moreover, we  
predict the $B_d\to\pi^0K_{\rm S}$ CP asymmetries as follows \cite{BFRS-III}:
\begin{equation}
{\cal A}_{\rm CP}^{\rm dir}(\pi^0 K_{\rm S})  %(B_d\to\pi^0 K_{\rm S})
=+0.05^{+0.24}_{-0.29},\quad 
{\cal A}_{\rm CP}^{\rm mix}(\pi^0 K_{\rm S})%(B_d\to\pi^0 K_{\rm S})
=-0.99^{+0.04}_{-0.01}.
\end{equation}
Recently, the BaBar collaboration reported the results of 
$0.40^{+0.27}_{-0.28}\pm0.10$ and $-0.48_{-0.38}^{+0.47}\pm0.11$ 
for these direct and mixing-induced CP asymmetries, respectively 
\cite{browder-talk}.

Let us finally note that we may complement the $B\to\pi\pi$ data in a
variety of ways with the experimental information provided by the 
$B_d\to\pi^\mp K^\pm$ modes, allowing us to determine $\gamma$ as 
well. If we take also the constraints from the whole 
$B\to \pi K$ system into account, we find results for $\gamma$ in remarkable 
agreement with the UT fits, i.e.\ we arrive at a very consistent overall 
picture \cite{BFRS-III}. In the future, 
$B_s\to K^+K^-$ will provide a powerful tool for the simultaneous 
determination of $\gamma$ and $(d,\theta)$ \cite{RF-BsKK}.

{\bf 4.}\
The implications of enhanced $Z^0$ penguins with a large new
complex phase for rare and CP-violating $K$ and $B$ decays
were already discussed in  
\cite{Buras:1998ed,Buras:1999da,Buchalla:2000sk}, where 
model-independent analyses and studies within particular supersymmetric 
scenarios were presented. Here we determine the size of the enhancement 
of the $Z^0$-penguin function $C$ and the magnitude of its complex phase 
through the $B\to\pi K$ data as discussed above. Performing a 
renormalization-group analysis as in \cite{BFRS-I} yields 
\be\label{RG}
C(\bar q)= 2.35~ \bar q e^{i\phi} -0.82,\quad 
\bar q= q \left[\frac{|V_{ub}/V_{cb}|}{0.086}\right].
\ee
Evaluating, in the spirit of \cite{Buras:1998ed,Buras:1999da,BFRS-I},
the relevant box-diagram contributions within the SM and using
(\ref{RG}), we can calculate the short-distance functions
\begin{equation}\label{X-C-rel}
X=C(\bar q)+0.73 \quad \mbox{and} \quad Y=C(\bar q)+0.18,
\end{equation}
which govern the rare $K$, $B$ decays with $\nu\bar\nu$ and $l^+l^-$ 
in the final states, respectively. 

The central value for $Y$ resulting from (\ref{q-det}) violates the 
upper bound $|Y|\le 2.2$ following from the BaBar and Belle data on 
$B\to X_s\mu^+\mu^-$ \cite{Kaneko:2002mr}, and the upper bound on
$\mbox{BR}(K_{\rm L}\to \pi^0 e^+e^-)$ of $2.8\times 10^{-10}$ from KTeV 
\cite{KTEVKL}. However, we may still encounter significant deviations from 
the SM. In order to illustrate this exciting feature, we consider
only the subset of those values of $(q,\phi)$ in (\ref{q-det}) that 
satisfy the constraint of $|Y|= 2.2$. If we then introduce the 
CP-violating weak 
phases $\theta_C$, $\theta_X$ and $\theta_Y$, and use (\ref{X-C-rel}) and 
(\ref{RG}), we obtain  
\be\label{CXY}
\begin{array}{rclcrcl}
|C|&=&2.24\pm 0.04,&&\theta_C&=&-(105\pm 12)^\circ,\\
|X|&=&2.17\pm0.12,&&\theta_X&=&-(87\pm12)^\circ,\\
|Y|&=&2.2\,\, {\rm (input)}, &&\theta_Y&=&-(103\pm12)^\circ.
\end{array}
\ee
This should be compared with the SM values $C=0.79$, $X=1.53$ and $Y=0.98$ 
for $m_t(m_t)=167~{\rm GeV}$.

The enhanced function $|C|$ and its large complex phase may affect the 
usual analysis of the UT \cite{CKM-Book} through double $Z^0$-penguin 
contributions to $\varepsilon_K$ and $\Delta M_{s,d}$, but as demonstrated 
in \cite{BFRS-III}, these effects can be neglected. Inserting then 
the values of $|X|e^{i\theta_X}$ and $|Y|e^{i\theta_Y}$ listed in 
(\ref{CXY}) into the known formulae for rare $K$- and $B$-decay 
branching ratios \cite{Schladming}, we obtain the following results:

a) 
For the very clean $K\to\pi\nu\bar\nu$ decays, we find
\begin{equation}\label{kpnr}
\begin{array}{rcl}
\mbox{BR}(\kpn)&=&(7.5\pm 2.1)\times 10^{-11}\\
\mbox{BR}(\klpn)&=&(3.1\pm 1.0)\times 10^{-10},
\end{array}
\end{equation}
to be compared with the SM estimates 
$(7.7 \pm 1.1)\times 10^{-11}$ and 
$(2.6 \pm 0.5)\times 10^{-11}$ \cite{Gino03},
respectively, and the AGS E787 result 
$\mbox{BR}(K^+ \rightarrow \pi^+ \nu \bar{\nu})=
(15.7^{+17.5}_{-8.2})\times 10^{-11}$ \cite{Adler97}. 
The enhancement of $\mbox{BR}(\klpn)$ by one order of magnitude 
and the pattern in (\ref{kpnr}) are dominantly the 
consequences of $\beta_X=\beta-\theta_X\approx
110^\circ$, as 
\be
\frac{\mbox{BR}(\klpn)}{\mbox{BR}(\klpn)_{\rm SM}}=
\left|\frac{X}{X_{\rm SM}}\right|^2
\left[\frac{\sin\beta_X}{\sin\beta}\right]^2
\ee
\be
\frac{\mbox{BR}(\klpn)}{\mbox{BR}(\kpn)}\approx 4.4\times (\sin\beta_X)^2
\approx (4.2\pm 0.2).  
\ee
Interestingly, the above ratio turns out to be very close to its absolute
upper bound in \cite{GRNIR}.
A spectacular 
implication of these findings is a strong violation of 
$(\sin 2 \beta)_{\pi \nu\bar\nu}=(\sin 2 \beta)_{\psi K_{\rm S}}$ 
\cite{BBSIN}, which is valid in the SM and any model with minimal
flavour violation. Indeed, we find
\be
(\sin 2 \beta)_{\pi \nu\bar\nu}=\sin 2\beta_X= -(0.69^{+0.23}_{-0.41}),
\ee
in striking disagreement with $(\sin 2 \beta)_{\psi K_{\rm S}}= 0.74\pm0.05$.
  
b) Another implication is the large branching ratio
\be
\mbox{BR}(K_{\rm L}\to\pi^0e^+e^-)= (7.8\pm 1.6)\times 10^{-11},
\ee
which is governed by direct CP violation. On the other hand, the
SM result $(3.2^{+1.2}_{-0.8})\times 10^{-11}$ \cite{BI03} is dominated 
by indirect CP violation. Moreover, the integrated forward--backward CP 
asymmetry for $B_d\to K^*\mu^+\mu^-$ \cite{Buchalla:2000sk}, which is 
given by
\be
A^{\rm CP}_{\rm FB}=(0.03\pm0.01)\times \tan\theta_Y, 
\ee
can be very large in view of $\theta_Y\approx -100^\circ$.

c)  
Next, $\mbox{BR}(B\to X_{s,d}\nu\bar\nu)$ and 
$\mbox{BR}(B_{s,d}\to \mu^+\mu^-)$ are enhanced by factors of $2$ and $5$, 
respectively, whereas the impact on $K_{\rm L}\to \mu^+\mu^-$ is rather 
moderate.

d) 
As emphasized in \cite{Buras:1998ed}, enhanced $Z^0$ penguins may play an 
important r\^ole in $\epe$. The enhanced value of $|C|$ and its large negative 
phase suggested by the $B\to\pi K$ analysis require a significant enhancement 
of the relevant hadronic matrix element of the QCD penguin operator $Q_6$, 
with respect to the one of the EW penguin operator $Q_8$, to be 
consistent with the $\epe$ data \cite{BFRS-III}. 

e) 
We have also explored the implications for the decay $B_d\to\phi K_{\rm S}$ 
\cite{BFRS-III}. Large hadronic uncertainties preclude a precise 
prediction, but assuming that the sign of the cosine 
of a strong phase agrees with factorization, we find that 
$(\sin 2 \beta)_{\phi K_{\rm S}}>(\sin 2 \beta)_{\psi K_{\rm S}}$,
where $(\sin 2 \beta)_{\phi K_{\rm S}}\sim1$ may well be possible.
This pattern is qualitatively different from the present $B$-factory 
data \cite{browder-talk}, which are, however, not yet conclusive. 
On the other hand, a future confirmation of this pattern would be 
another signal of enhanced CP-violating $Z^0$ penguins at work. 

In the next couple of years, it will be very exciting to follow the
development of the values of the observables considered in this letter
and to monitor them by using the strategies presented here and in 
\cite{BFRS-III}.

\vspace*{1mm}
%******************************************************************************
{\bf Acknowledgements}
%******************************************************************************
This research was partially supported by the German BMBF, contract 05HT1WOA3.

\end{document}